\pdfoutput=1
\documentclass[11pt,a4paper]{article}

\usepackage[T1]{fontenc}
\usepackage{lmodern}
\usepackage{microtype}
\usepackage[margin=2.7cm]{geometry}
\usepackage{amsmath,amssymb,amsthm}
\usepackage{booktabs}
\usepackage{array}
\usepackage{xspace}
\usepackage{xcolor}
\usepackage{tikz}
\usepackage{enumitem}
\usepackage{cite}
\usepackage{authblk}
\usepackage[hidelinks]{hyperref}
\usepackage{xurl}
\usepackage{orcidlink}%

\makeatletter\@clubpenalty=\clubpenalty\makeatother

\usetikzlibrary{arrows,automata,positioning}
\tikzset{>=stealth}
\tikzset{every edge/.append style={shorten >=1pt}}
\tikzset{
  game/.style={node distance=1.3cm and 1.5cm, on grid, auto, semithick,
               initial text=, initial distance=6mm,
               every state/.style={minimum size=6.5mm, inner sep=1pt, font=\small}},
  eve/.style={state},
  adam/.style={state, rectangle},
  sink/.style={state, draw=black!60},
  c1/.style={fill=orange!45},
  c2/.style={fill=cyan!40},
  c3/.style={fill=green!45},
  c4/.style={fill=violet!35},
  c5/.style={fill=red!30},
  c6/.style={fill=yellow!60},
}

\theoremstyle{plain}
\newtheorem{theorem}{Theorem}[section]
\newtheorem{lemma}[theorem]{Lemma}

\theoremstyle{definition}
\newtheorem{definition}[theorem]{Definition}
\newtheorem{example}[theorem]{Example}
\theoremstyle{remark}
\newtheorem{remark}[theorem]{Remark}

\newcommand{\Eve}{Eve\xspace}
\newcommand{\Adam}{Adam\xspace}
\newcommand{\VE}{V_{\mathrm{E}}}
\newcommand{\VA}{V_{\mathrm{A}}}
\newcommand{\game}{\mathcal{G}}
\newcommand{\arena}{\mathcal{A}}
\newcommand{\mem}{\mathcal{M}}
\newcommand{\upd}{\mu}
\newcommand{\nxt}{\nu}
\newcommand{\sink}{\bot}
\newcommand{\col}{\operatorname{col}}
\newcommand{\Succ}{\operatorname{Succ}}
\newcommand{\memof}{\operatorname{mem}}
\newcommand{\set}[1]{\{#1\}}
\newcommand{\setk}{[k]}
\newcommand{\opt}[1]{\neg #1}

\title{Lower Bounds on Adam's Memory in Generalized Reachability Games with Target Sets of Size Two}
\author[1,2]{Daniel Neider\,\orcidlink{0000-0001-9276-6342}}
\affil[1]{TU Dortmund University, Dortmund, Germany}
\affil[2]{Research Center for Trustworthy AI, UA Ruhr, Dortmund, Germany}
\date{}

\begin{document}

\maketitle

\begin{abstract}
In a generalized reachability game, \Eve tries to visit every one of $k$ target sets and \Adam tries to prevent at least one visit.
When every target set has at most two vertices, the complexity of deciding the winner is open.
Fijalkow and Horn, and later Bose et al., asked whether four memory states suffice whenever \Adam wins, which would place the problem in $\mathrm{coNP}^{\mathrm{NP}}$.
Moreover, Fijalkow conjectured that two states suffice on acyclic arenas.
We answer both questions in the negative.
We give an acyclic game with five target sets in which \Adam wins but needs three memory states.
Furthermore, we give two cyclic games with five and six target sets in which \Adam needs five and six memory states, respectively.
The latter two games extend the example of Löding that provides the known lower bound of four memory states.
\end{abstract}

\section{Introduction}
\label{sec:intro}

In a \emph{generalized reachability game}, two players, \Eve and \Adam, move a token through a finite directed graph whose vertices are partitioned between them.
The game is equipped with target sets~$F_1, \dots, F_k$ of vertices, and \Eve wins a play if it visits every target set; conversely,  \Adam wins if he can prevent at least one visit.
Fijalkow and Horn studied these games and showed that deciding the winner is PSPACE-complete in general~\cite{DBLP:journals/corr/abs-1010-2420,FijalkowHorn2013}.\footnote{The same result appears earlier for LTL games built from eventualities and conjunctions~\cite{DBLP:journals/tocl/AlurT04}.}
Their surprising finding is that the complexity depends on the size of the target sets: with target sets of size one the problem is in P, with target sets of size three it is already PSPACE-complete, and with target sets of size at most two it is in PSPACE, but its exact complexity is open.
The extended version~\cite{DBLP:journals/corr/abs-2607-14199} of the paper of Bose et al.~\cite{DBLP:journals/corr/abs-2509-14091} calls this ``the major remaining open problem''.

One way to improve the PSPACE upper bound rests on \Adam's memory.
If a fixed number of memory states suffices whenever \Adam wins, the problem is in $\mathrm{coNP}^{\mathrm{NP}}$: guess a winning strategy of \Adam with so many states, compose it with the game and solve the resulting one-player game.
In this context, Fijalkow and Horn~\cite{DBLP:journals/corr/abs-1010-2420} and Bose et al.~\cite{DBLP:journals/corr/abs-2607-14199} ask whether four memory states suffice, which both papers call plausible.
The number four comes from an example with four target sets due to Löding, in which \Adam needs four states, and both papers remark that they ``could not generalize this example to obtain a better lower bound than~4''.
Fijalkow further conjectured in a blog post that two states suffice when the graph is acyclic~\cite{FijalkowBlog2018}.
The general bound known is $\binom{k}{\lfloor k/2 \rfloor}$ memory states for $k$ target sets of any size, and it is tight for target sets of unbounded size~\cite{DBLP:journals/corr/abs-1010-2420,DBLP:journals/corr/abs-2607-14199,DBLP:conf/fsttcs/ColcombetFH14}.

\paragraph{Contributions.}
We refute both conjectures, the one for acyclic arenas with one game and the four-state bound with two.
First, Section~\ref{sec:three} provides an acyclic game~$\game_3$ with five target sets of size two in which \Adam wins but needs three memory states; here and below, the index of a game is the number of memory states \Adam needs in it.
Padding by further colors that every play visits gives such a game for every $k \geq 5$ (Section~\ref{sec:five-or-more}).
Consequently, the acyclic bound of two fails from five target sets on.

Second, Section~\ref{sec:cyclic} shows that Löding's example does generalize.
It gives two games, $\game_5$ and $\game_6$, with five and six target sets, in which \Adam needs five and six memory states, respectively.
Both games have the shape of Löding's example: an acyclic part in which only \Eve moves, then a single vertex~$c$ of \Adam with one move per color, each move leading back to~$c$.
The change that makes them work is small: some of \Adam's moves at $c$ lead back to the first of the two vertices of a color, where in Löding's example every move leads back to a second one.
With the same padding, these games refute the four-state bound for every $k \geq 5$ and raise the lower bound to six for every $k \geq 6$ (Section~\ref{sec:five-or-more}).

\paragraph{How the results were obtained.}
The three games were found with computer help.
The game~$\game_3$ turned up in a random walk over acyclic games with five colors, which mutated known games and cut each mutant down to a subgame that \Adam still wins but in which no edge of his can be removed; we reduced the smallest game found there, with 22 vertices, to the plays that its lower bound uses.
The game~$\game_5$ came from an exhaustive enumeration of systems of \Eve's paths to \Adam's vertex in the shape of Löding's example; we added the options to one of them and removed redundant edges.
The game~$\game_6$ came from a SAT encoding of this shape, asked for a model without redundant edges.
That each game needs the stated memory was then confirmed by at least two SAT solvers, and the strategies they found were replayed by an independent checker.
The solvers checked a SAT encoding in the style of bounded synthesis~\cite{DBLP:journals/sttt/FinkbeinerS13}, which fixes the number of memory states and asks for a winning strategy of \Adam with this number of states.
However, the proofs in this note are pen-and-paper proofs, not machine-checked ones.

\paragraph{Related work.}
Memory requirements of games on graphs are surveyed in~\cite{fijalkow2025gamesgraphslogicautomata,DBLP:conf/fsttcs/BrihayeGMR23}.
\Adam's objective here is a safety objective, and the general theory of memory for safety objectives~\cite{DBLP:conf/fsttcs/ColcombetFH14,DBLP:journals/lmcs/ColcombetFH24} explains the antichain bound $\binom{k}{\lfloor k/2 \rfloor}$; it uses neither the size of the target sets nor the shape of the arena.
In a setting where colors label the edges, memory that reads the edges of a play has been compared with memory that reads only their colors~\cite{DBLP:conf/fsttcs/BouyerRV22}.
Our lower bounds hold for both.

\section{Preliminaries}
\label{sec:prelims}

We assume familiarity with games on graphs~\cite{DBLP:conf/dagstuhl/2001automata,fijalkow2025gamesgraphslogicautomata} and recall the notions we need, in the terminology of~\cite{DBLP:journals/corr/abs-2607-14199}.
For $k \in \mathbb{N}$, we write $\setk = \set{1, \dots, k}$.

\subsection{Arenas, Games and Strategies}
\label{sec:games}

A game is played by two players, whom we call \Eve and \Adam, who move a token through a finite graph, called \emph{arena}.
Each vertex of the graph belongs to one of the players, and the owner of the vertex the token is on chooses the edge along which the token moves next.
Since we want plays to go on forever, every vertex has at least one outgoing edge.
Formally, we define arenas as follows.

\begin{definition}[Arena]
An \emph{arena} is a four-tuple $\arena = (V, \VE, \VA, E)$ consisting of a finite set~$V$ of vertices, a partition of $V$ into the vertices~$\VE$ of \Eve and the vertices~$\VA$ of \Adam, and a set~$E \subseteq V \times V$ of edges such that every vertex has at least one outgoing edge.
\end{definition}

We write $u \to w$ for $(u, w) \in E$ and $\Succ(u) = \set{w \mid u \to w}$ for the set of successors of $u$.
A \emph{path} is a finite sequence~$v_0 v_1 \cdots v_j$ of vertices with $v_i \to v_{i+1}$ for every $i < j$; it \emph{starts} in $v_0$ and \emph{ends} in $v_j$.
If a path~$\pi$ ends in the vertex in which a path~$\rho$ starts, then $\pi\rho$ denotes the path that follows $\pi$ and then $\rho$, with the common vertex written once.
A \emph{play} is an infinite sequence~$v_0 v_1 v_2 \cdots$ of vertices with $v_i \to v_{i+1}$ for every $i \geq 0$.
We say that a path or play \emph{visits} the vertices that occur in it.
A vertex of \Adam with at least two successors is called an \emph{\Adam choice vertex}.
A vertex with exactly one successor is \emph{forced}.
In drawings, circles belong to \Eve and squares to \Adam.

A \emph{sink} is a vertex~$v$ with $\Succ(v) = \set{v}$.
An arena is \emph{acyclic} if its only cycles are the self-loops at sinks; otherwise we call it \emph{cyclic}.
In an acyclic arena, every play reaches a sink and stays there, meaning it visits every vertex other than that sink at most once.

A game consists of an arena, an initial vertex where the token starts, and a winning condition that says which plays \Eve wins.
We study games in which \Eve has to visit every one of several target sets of vertices.
Formally, these games are defined as follows.

\begin{definition}[Generalized reachability game]
\label{def:game}
A \emph{generalized reachability game} is a tuple $\game = (\arena, s, F_1, \dots, F_k)$ consisting of an arena~$\arena$, an initial vertex~$s \in V$ and \emph{target sets} $F_1, \dots, F_k \subseteq V$.
\Eve wins a play that starts in $s$ if it visits every target set; otherwise \Adam wins it.
\end{definition}

Let $\game = (\arena, s, F_1, \dots, F_k)$ be a generalized reachability game.
Its \emph{plays} are the plays that start in~$s$, and its \emph{finite plays} are the paths that start in~$s$.
We call $i \in \setk$ a \emph{color}, say that a vertex~$v \in F_i$ \emph{has color~$i$}, and write $\col(v) = \set{i \in \setk \mid v \in F_i}$ for the set of colors of $v$.
Furthermore, we say that a path or play \emph{visits color~$i$} if it visits a vertex of color~$i$.
For a path~$\pi$ through the arena, $\col(\pi)$ is the union of the colors of its vertices.

Throughout this note, every target set has at most two vertices, so every color has at most two vertices (i.e., $|F_i| \leq 2$ for all $i \in \setk$).
By a standard construction (split a vertex with several colors into a chain of vertices with one color each) we may and do assume that the target sets are pairwise disjoint, so that every vertex has at most one color.
This change neither affects the winner nor the number of memory states \Adam needs: the new vertices have one successor each, and an update along the edge into the split vertex can be made along the chain instead, and conversely.

The winner of a game is defined by means of strategies.
In the following, let $P \in \set{\mathrm{E}, \mathrm{A}}$ be a player, where $\mathrm{E}$ stands for \Eve and $\mathrm{A}$ for \Adam.

\begin{definition}[Strategy]
\label{def:strategy}
A \emph{strategy} of $P$ is a function~$\sigma \colon V^* V_P \to V$ such that $\sigma(wv) \in \Succ(v)$ for all $w \in V^*$ and $v \in V_P$.
\end{definition}

A play~$v_0 v_1 v_2 \cdots$ of $\game$ is \emph{consistent} with $\sigma$ if $v_{j+1} = \sigma(v_0 \cdots v_j)$ for every $j \geq 0$ with $v_j \in V_P$.
Likewise, a finite play~$v_0 \cdots v_n$ of $\game$ is \emph{consistent} with $\sigma$ if $v_{j+1} = \sigma(v_0 \cdots v_j)$ for every $j < n$ with $v_j \in V_P$.

The strategy~$\sigma$ is \emph{winning} if $P$ wins every play consistent with it.
A player \emph{wins the game} if they have a winning strategy.
Since every vertex has a successor, every finite play consistent with $\sigma$ is a prefix of a play consistent with $\sigma$.
Hence, no finite play consistent with a winning strategy of \Adam visits every color.
Generalized reachability games are determined, that is, in every game exactly one player wins, since the winner of a play depends only on the set of colors it visits.
Recording this set in the vertices yields a reachability game on a finite arena~\cite{DBLP:journals/corr/abs-1010-2420,DBLP:journals/corr/abs-2607-14199}, and reachability games are determined~\cite{DBLP:journals/tcs/Zielonka98,DBLP:conf/dagstuhl/2001automata}.

Our lower bounds rest on the moves that a winning strategy of Adam can make after a given finite play.
Since Eve wins a play if and only if it visits every color, these moves depend only on the last vertex of the finite play and the set of colors visited so far.
We call such moves \emph{good} and define them next.

\begin{definition}[Good move]
\label{def:good}
Let $u$ be an \Adam vertex and $S \subseteq \setk$ a set of colors.
A successor~$w$ of $u$ is a \emph{good move at $S$} if \Adam wins the generalized reachability game~$(\arena, w, (F_i)_{i \in \setk \setminus S})$, that is, the game that starts at $w$ and in which only the colors outside $S$ remain to be visited.
\end{definition}

\subsection{Memory}

In general, a strategy as defined above is an infinite object.
However, we are interested in strategies that can be implemented by a finite automaton that reads the play and whose state determines the next move: the automaton is the \emph{memory} of the strategy, and we ask how many states it needs.

\begin{definition}[Memory structure~\cite{DBLP:conf/lics/DziembowskiJW97}]
\label{def:memory}
A \emph{memory structure} $\mem = (M, q_0, \upd)$ consists of a finite set~$M$ of \emph{memory states}, an initial state~$q_0 \in M$ and an update function~$\upd \colon M \times E \to M$.
The \emph{run} of $\mem$ from a memory state~$q$ along a path~$v_0 v_1 \cdots v_j$ ends in the memory state~$\upd^*(q, v_0 \cdots v_j)$, defined by $\upd^*(q, v_0) = q$ and $\upd^*(q, v_0 \cdots v_i) = \upd(\upd^*(q, v_0 \cdots v_{i-1}), (v_{i-1}, v_i))$ for $1 \leq i \leq j$.
\end{definition}

For a finite play~$\pi$ we write $\upd^*(\pi) = \upd^*(q_0, \pi)$ for the memory state \emph{at} its last vertex; it includes the update along the edge into that vertex.
Since the update is applied edge by edge, $\upd^*(q, \pi\rho) = \upd^*(\upd^*(q, \pi), \rho)$ whenever $\pi$ ends where $\rho$ starts.

A memory structure only keeps track of the play.
A strategy with memory chooses \Adam's successor from the last vertex of the finite play and the memory state at it.

\begin{definition}[Strategy with memory]
A \emph{strategy of \Adam with memory~$\mem$} is a pair~$\sigma = (\mem, \nxt)$, where $\mem$ is a memory structure and $\nxt \colon \VA \times M \to V$ with $\nxt(u, q) \in \Succ(u)$ is a \emph{next-move function}.
\end{definition}

Every strategy~$\sigma = (\mem, \nxt)$ of \Adam with memory implements a strategy of \Adam in the sense of Definition~\ref{def:strategy} by mapping every finite play~$wu$ with $u \in \VA$ to $\nxt(u, \upd^*(wu))$ (and every other word~$wu \in V^* \VA$ to a fixed successor of~$u$).
\Adam \emph{wins with $m$ memory states} if he has a winning strategy~$(\mem, \nxt)$ with $|M| = m$, and a strategy with one memory state is called \emph{positional}.
If \Adam wins $\game$, let $\memof(\game)$ denote the least $m$ such that he wins with $m$ states.
Some authors use other memory models, in which the memory reads only the vertices of a play, or only colors, which label either its vertices or its edges~\cite{DBLP:conf/fsttcs/BouyerRV22}.
Our model can simulate both with the same number of memory states.
For a memory that reads only vertices, a memory structure with the same memory states uses an update~$\upd(q, (u, v))$ that depends only on $q$ and~$v$.
For a memory that reads only colors, the update depends only on $q$ and the colors of~$v$, or of the edge~$(u, v)$ if colors label edges.
If the memory also reads the initial vertex, the state after this reading serves as the initial state~$q_0$, since the initial vertex is fixed.
By this simulation, every lower bound that we prove also holds in these models.

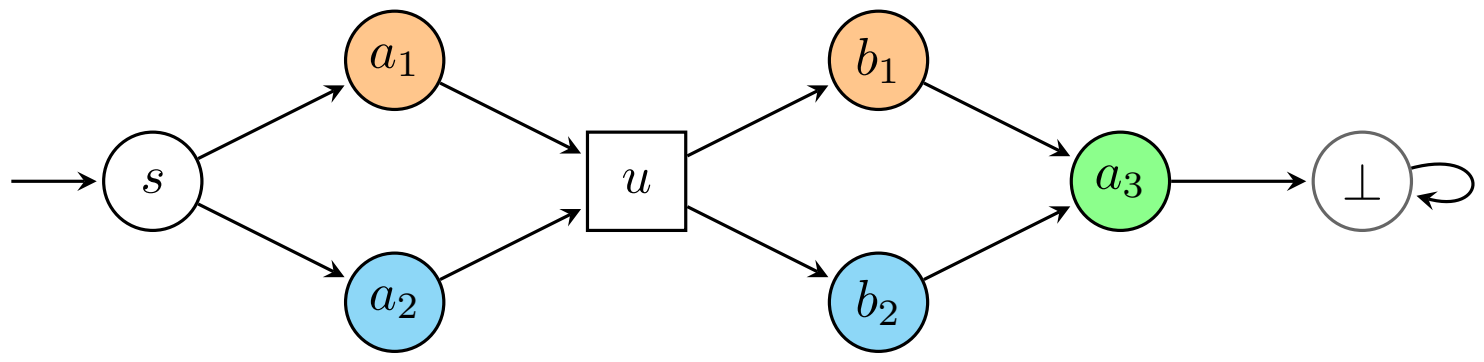
\begin{figure}[htb]
\centering
\begin{tikzpicture}[game]
  \node[eve, initial] (s)  at (0, 0)   {$s$};
  \node[eve, c1]   (a1) at (1.6, 0.8)  {$a_1$};
  \node[eve, c2]   (a2) at (1.6, -0.8) {$a_2$};
  \node[adam]      (u)  at (3.2, 0)    {$u$};
  \node[eve, c1]   (b1) at (4.8, 0.8)  {$b_1$};
  \node[eve, c2]   (b2) at (4.8, -0.8) {$b_2$};
  \node[eve, c3]   (a3) at (6.4, 0)    {$a_3$};
  \node[sink]      (t)  at (8.0, 0)    {$\sink$};
  \path[->]
    (s) edge (a1) (s) edge (a2)
    (a1) edge (u) (a2) edge (u)
    (u) edge (b1) (u) edge (b2)
    (b1) edge (a3) (b2) edge (a3)
    (a3) edge (t)
    (t) edge [loop right] (t);
\end{tikzpicture}
\caption{The game of Example~\ref{ex:pair-gadget}, with target sets~$F_1 = \set{a_1, b_1}$, $F_2 = \set{a_2, b_2}$ and $F_3 = \set{a_3}$.
Colored vertices are filled, one fill per color. The initial vertex is marked by an incoming arrow.}
\label{fig:pair-gadget}
\end{figure}

\begin{example}
\label{ex:pair-gadget}
Consider the game~$\game$ of Figure~\ref{fig:pair-gadget}.
This game is acyclic and has three colors, with the target sets~$F_1 = \set{a_1, b_1}$, $F_2 = \set{a_2, b_2}$ and $F_3 = \set{a_3}$.
\Eve chooses at vertex~$s$ whether to visit vertices~$a_1$ or $a_2$, \Adam chooses at vertex~$u$ whether to visit vertices~$b_1$ or $b_2$, and then vertex~$a_3$ and the sink~$\sink$ are visited.
Every play visits color~$3$, so \Eve wins a play if and only if it visits both colors~$1$ and $2$, that is, if and only if \Adam's choice at $u$ is the color that \Eve avoided.
\Adam wins by repeating at $u$ the color \Eve chose.
For this, two memory states suffice, since the bit ``has $a_1$ been visited?'' tells \Adam what to do at $u$.
One memory state does not suffice, because a positional strategy of \Adam fixes one move at $u$, say $b_1$, and \Eve wins against it by visiting $a_2$.
Hence, $\memof(\game) = 2$.
\end{example}

\section{Three Memory States on an Acyclic Arena}
\label{sec:three}

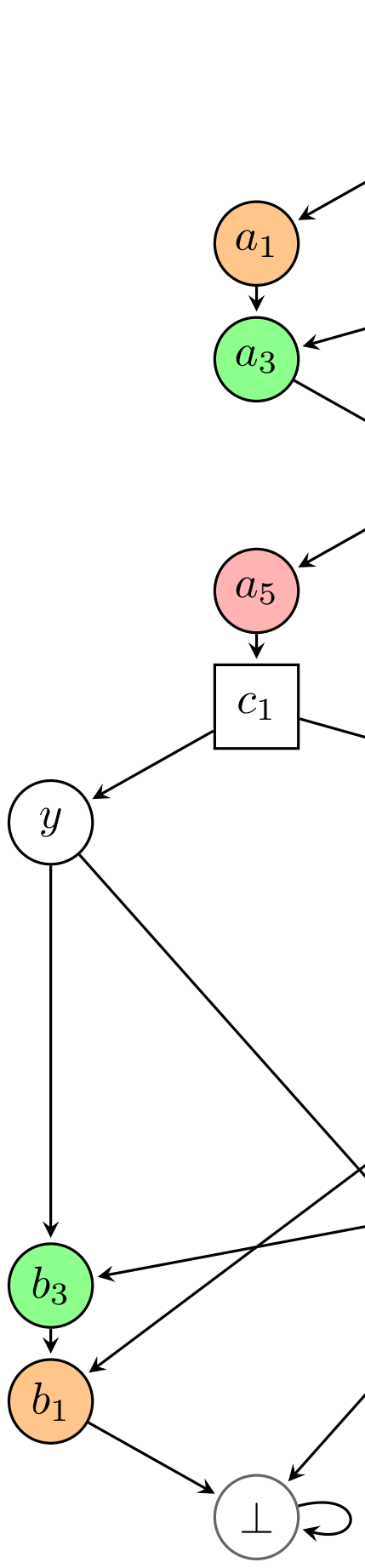
\begin{figure}[t]
\centering
\begin{tikzpicture}[game, x=1.6cm, y=0.9cm]
  \node[eve, initial above] (s) at (2, 0) {$s$};
  \node[eve, c1]  (a1) at (1, -1)  {$a_1$};
  \node[eve, c2]  (a2) at (3, -1)  {$a_2$};
  \node[eve, c3]  (a3) at (1, -2)  {$a_3$};
  \node[eve, c4]  (a4) at (3, -2)  {$a_4$};
  \node[eve]      (x)  at (2, -3)  {$x$};
  \node[eve, c5]  (a5) at (1, -4)  {$a_5$};
  \node[adam]     (u1) at (1, -5)  {$c_1$};
  \node[eve]      (y)  at (0, -6)  {$y$};
  \node[eve, c4]  (b4) at (3, -6)  {$b_4$};
  \node[eve]      (z)  at (3, -7)  {$z$};
  \node[eve, c5]  (b5) at (3, -8)  {$b_5$};
  \node[adam]     (u2) at (3, -9)  {$c_2$};
  \node[eve, c3]  (b3) at (0, -10) {$b_3$};
  \node[eve, c2]  (b2) at (2, -10) {$b_2$};
  \node[eve, c1]  (b1) at (0, -11) {$b_1$};
  \node[sink]     (bot) at (1, -12) {$\sink$};
  \path[->]
    (s) edge (a1) (s) edge (a2)
    (a1) edge (a3) (a2) edge (a3) (a2) edge (a4)
    (a3) edge (x) (a4) edge (x)
    (x) edge (a5) (x) edge (b4)
    (a5) edge (u1)
    (u1) edge (y) (u1) edge (b4)
    (y) edge (b3) (y) edge (b2)
    (b4) edge (z)
    (z) edge (b5) (z) edge (b1) (z) edge [bend right=15] (b2)
    (b5) edge (u2)
    (u2) edge (b3) (u2) edge (b2)
    (b3) edge (b1)
    (b1) edge (bot) (b2) edge (bot)
    (bot) edge [loop right] (bot);
\end{tikzpicture}
\caption{The game~$\game_3$ of Example~\ref{ex:m3} ($k = 5$, acyclic), with the target sets~$F_j = \set{a_j, b_j}$ for $j = 1, \dots, 5$.}
\label{fig:m3}
\end{figure}

Fijalkow conjectured that \Adam wins every acyclic game with target sets of size at most two that he wins at all with two memory states~\cite{FijalkowBlog2018}.
The game~$\game_3$ of this section refutes the conjecture for $k = 5$, and by a padding argument for every $k \geq 5$ (Section~\ref{sec:five-or-more}).
The idea is simple: \Eve can choose among three paths before \Adam has a choice, and his memory must tell all three apart.
The game~$\game_3$ is defined in Example~\ref{ex:m3} and illustrated in Figure~\ref{fig:m3}.

\begin{example}[\Adam needs three memory states; $k = 5$, acyclic]
\label{ex:m3}
Figure~\ref{fig:m3} shows the acyclic game~$\game_3$ with initial vertex~$s$ and sink~$\sink$.
\Adam owns $c_1$ and $c_2$, and \Eve owns all other vertices.
The target sets are $F_j = \set{a_j, b_j}$ for $j = 1, \dots, 5$.
The edges of $\game_3$ are drawn as arrows in Figure~\ref{fig:m3}.
\end{example}

The game proceeds in three stages.
First, \Eve chooses one of three paths from $s$ to $x$, namely $\pi_1 = s\,a_1\,a_3\,x$, $\pi_2 = s\,a_2\,a_3\,x$ and $\pi_3 = s\,a_2\,a_4\,x$.
Each of them visits two of the colors~$1, \dots, 4$ and leaves the other two unvisited:
\begin{center}
\begin{tabular}{@{}lll@{}}
\toprule
path & colors visited & colors among $1, \dots, 4$ not visited \\
\midrule
$\pi_1$ & $1, 3$ & $2, 4$ \\
$\pi_2$ & $2, 3$ & $1, 4$ \\
$\pi_3$ & $2, 4$ & $1, 3$ \\
\bottomrule
\end{tabular}
\end{center}

Second, \Eve chooses at $x$ which of \Adam's two choice vertices comes next.
The vertices of color~$5$ are $a_5$ and $b_5$; both are forced, and their successors are $c_1$ and $c_2$.
Since \Eve has to visit color~$5$ to win, the play has to pass $c_1$ or $c_2$, and at $x$ she decides which comes first: the move to $a_5$ leads to $c_1$, and after the move to $b_4$ she can continue through $z$ and $b_5$ to $c_2$, visiting color~$4$ on the way.

Third, \Adam moves at the choice vertex that \Eve has chosen.
To win the game, \Adam has to keep \Eve from one of the colors not visited so far.
He can do so as follows:
\begin{itemize}
\item \emph{At $c_1$} \Adam chooses between $y$ and $b_4$.
The move to $b_4$ visits color~$4$, whereas no path from $y$ visits color~$4$.
So \Adam moves to $y$ if color~$4$ has not been visited (paths~$\pi_1$ and~$\pi_2$), and to $b_4$ if it has (path~$\pi_3$).
The other move loses in both cases:
if color~$4$ has not been visited and \Adam moves to $b_4$, then one color is left (color~$2$ after $\pi_1$, color~$1$ after $\pi_2$), and \Eve visits it from $z$;
if color~$4$ has been visited and \Adam moves to $y$, then \Eve continues by $b_3\,b_1$ and visits the remaining colors~$3$ and~$1$.
\item \emph{At $c_2$} the colors~$4$ and~$5$ have been visited, and \Adam chooses between $b_2$ and $b_3$.
The move to $b_2$ visits color~$2$, and the move to $b_3$, whose only successor is $b_1$, visits the colors~$3$ and~$1$.
So \Adam moves to $b_3$ if color~$2$ has not been visited (path~$\pi_1$), and to $b_2$ if it has (paths~$\pi_2$ and~$\pi_3$).
The other move loses in both cases, since it visits every color not visited so far.
\end{itemize}
So each of \Adam's two moves depends on one fact about \Eve's path to~$x$:
\begin{center}
\begin{tabular}{@{}lllll@{}}
\toprule
path & color~$4$ visited? & move at $c_1$ & color~$2$ visited? & move at $c_2$ \\
\midrule
$\pi_1$ & no  & $y$   & no  & $b_3$ \\
$\pi_2$ & no  & $y$   & yes & $b_2$ \\
$\pi_3$ & yes & $b_4$ & yes & $b_2$ \\
\bottomrule
\end{tabular}
\end{center}

Any two rows of this table differ in the move at $c_1$ or in the move at $c_2$, which is why \Adam needs three memory states.
Take $\pi_2$ and $\pi_3$, and suppose that the memory is in the same state at $x$ after both.
If \Eve continues from $x$ to $c_1$, then the memory reads the same edges in both cases, so it is also in the same state at $c_1$, and \Adam makes the same move there.
However, \Adam has to move to $y$ after $\pi_2$ and to $b_4$ after $\pi_3$, so he loses in one of the two cases.
The same argument applies to $\pi_1$ and $\pi_2$ at $c_2$, and to $\pi_1$ and $\pi_3$ at either choice vertex.
Hence, for \Adam to win, the memory states at $x$ after the three paths must be pairwise different.

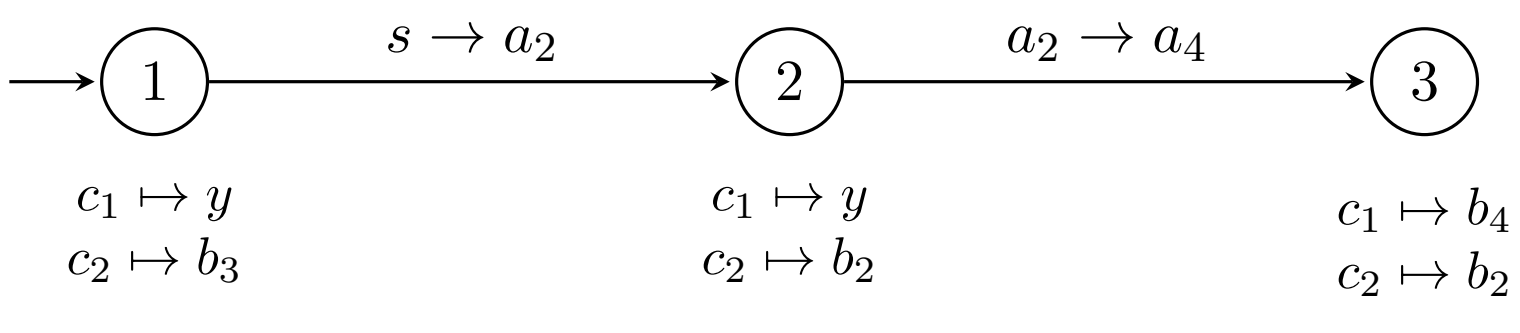
\begin{figure}[ht]
\centering
\begin{tikzpicture}[->, semithick, initial text=, initial distance=6mm,
                    every state/.style={minimum size=7mm, inner sep=1pt}]
  \node[state, initial] (q1) at (-4.2, 0) {$1$};
  \node[state]          (q2) at (0, 0)    {$2$};
  \node[state]          (q3) at (4.2, 0)  {$3$};
  \path (q1) edge node[above] {$s \to a_2$}   (q2)
        (q2) edge node[above] {$a_2 \to a_4$} (q3);
  \node[below=2mm of q1, align=center, font=\small] {$c_1 \mapsto y$\\ $c_2 \mapsto b_3$};
  \node[below=2mm of q2, align=center, font=\small] {$c_1 \mapsto y$\\ $c_2 \mapsto b_2$};
  \node[below=2mm of q3, align=center, font=\small] {$c_1 \mapsto b_4$\\ $c_2 \mapsto b_2$};
\end{tikzpicture}
\caption{A winning strategy of \Adam in $\game_3$ with three memory states.
An arrow labeled $u \to v$ is a memory update along the edge~$u \to v$ of $\game_3$.
In every other case, the memory keeps its state.
Below each state is \Adam's next-move function at $c_1$ and $c_2$ in that state.}
\label{fig:m3-strategy}
\end{figure}

\begin{theorem}
\label{thm:m3}
\Adam wins $\game_3$ with three memory states, and every winning strategy of \Adam has at least three (i.e., $\memof(\game_3) = 3$).
\end{theorem}

\begin{proof}[Proof of Theorem~\ref{thm:m3}]
We prove the two claims of the theorem in turn.
First, we give a strategy of \Adam with three memory states and show that it is winning.
Second, we show that every winning strategy of \Adam requires at least three memory states.

\medskip
\noindent\textit{Three states suffice.}
Figure~\ref{fig:m3-strategy} shows a winning strategy of \Adam with three memory states.
We first observe that its memory records the path to $x$.
Path~$\pi_1$ passes neither $a_2$ nor $a_4$ (the memory stays in state~$1$), path~$\pi_2$ passes $a_2$ and not $a_4$ (the memory moves to state~$2$), and path~$\pi_3$ passes $a_2$ and then $a_4$ (the memory moves to state~$2$ and then state~$3$).
Since no later edge changes the memory, the memory state at both $c_1$ and $c_2$ is the index~$i$ of the path~$\pi_i$ that the play followed.

Now consider a play that is consistent with this strategy, and let $i \in \set{1, 2, 3}$ be the index of the path $\pi_i$ that is taken to $x$.
Note that every path from $x$ to the sink visits $y$ or $z$.
We now analyze three possible outcomes and show that in each case, one color is not visited:
\begin{itemize}
  \item \emph{The play visits $y$.}
  Then, $i \in \set{1, 2}$ because \Adam moves from $c_1$ to $y$ only in states~$1$ and~$2$.
  Hence, the play does not visit $a_4$, which lies on neither $\pi_1$ nor $\pi_2$.
  It does not visit $b_4$ either, because there is no path from $b_4$ to $y$ and no path from $y$ to $b_4$.
  So, color~$4$ is not visited.
  \item \emph{The play visits $z$, and \Eve moves from $z$ to $b_1$ or $b_2$.}
  Then, the play visits neither $b_3$ nor $b_5$, because neither of them has a path to $z$ and neither is reachable from $b_1$ or $b_2$.
  If $i \in \set{1, 2}$, then the play does not visit $a_5$ either: the only successor of $a_5$ is $c_1$, where \Adam would move to $y$ in states~$1$ and~$2$, and there is no path from $y$ to $z$.
  So, color~$5$ is not visited.
  If $i = 3$, then the play does not visit $a_3$ either, which does not lie on $\pi_3$.
  Thus, color~$3$ is not visited.
  \item \emph{The play visits $z$, and \Eve moves from $z$ to $b_5$.}
  Then, the play visits $c_2$, which is the only successor of $b_5$.
  If $i = 1$, then \Adam moves from $c_2$ to $b_3$, and the play does not visit $b_2$ because $b_2$ has no path to $c_2$ and is not reachable from $b_3$.
  It does not visit $a_2$ either, which does not lie on $\pi_1$.
  So, color~$2$ is not visited.
  If $i \in \set{2, 3}$, then \Adam moves from $c_2$ to $b_2$, and the play does not visit $b_1$ because $b_1$ has no path to $c_2$ and is not reachable from $b_2$.
  It does not visit $a_1$ either, which lies on neither $\pi_2$ nor $\pi_3$.
  Thus, color~$1$ is not visited.
\end{itemize}
In conclusion, no play that is consistent with the strategy visits all five colors.
Thus, the strategy of Figure~\ref{fig:m3-strategy} is winning.

\medskip
\noindent\textit{Two states do not suffice.}
Let $\rho_1 = x\,a_5\,c_1$ and $\rho_2 = x\,b_4\,z\,b_5\,c_2$ be two paths from $x$ that end at the choice vertices~$c_1$ and $c_2$.
The paths~$\pi_i\rho_1$ and $\pi_i\rho_2$ contain no \Adam vertex before their last vertex.
Hence, the finite plays~$\pi_i\rho_1$ and $\pi_i\rho_2$ for $i \in \set{1, 2, 3}$ are consistent with every strategy of \Adam.

After $\pi_i\rho_1$ the play is at $c_1$ with the colors of $\pi_i$ and $5$; after $\pi_i\rho_2$ it is at $c_2$ with the colors of $\pi_i$, $4$ and $5$.
In each case, one of \Adam's two moves is \emph{losing}: after it, \Eve can ensure that the play visits all five colors.
We call the other move \emph{safe}.
For each of the six finite plays, the following table shows which move is losing and how \Eve then ensures all five colors:
\begin{center}
\begin{tabular}{@{}lllll@{}}
\toprule
finite play & choice vertex & colors & losing move and \Eve's reply & safe move \\
\midrule
$\pi_1\rho_1$ & $c_1$ & $1\,3\,5$ & $b_4$, then $z \to b_2$ (adds $4, 2$) & $y$ \\
$\pi_2\rho_1$ & $c_1$ & $2\,3\,5$ & $b_4$, then $z \to b_1$ (adds $4, 1$) & $y$ \\
$\pi_3\rho_1$ & $c_1$ & $2\,4\,5$ & $y$, then $y \to b_3 \to b_1$ (adds $3, 1$) & $b_4$ \\
$\pi_1\rho_2$ & $c_2$ & $1\,3\,4\,5$ & $b_2$ (adds $2$) & $b_3$ \\
$\pi_2\rho_2$ & $c_2$ & $2\,3\,4\,5$ & $b_3 \to b_1$ (adds $1$) & $b_2$ \\
$\pi_3\rho_2$ & $c_2$ & $2\,4\,5$ & $b_3 \to b_1$ (adds $3, 1$) & $b_2$ \\
\bottomrule
\end{tabular}
\end{center}

Now let $\sigma = (\mem, \nxt)$ with $\mem = (M, q_0, \upd)$ be a winning strategy of \Adam.
As argued above, each of the six finite plays is consistent with $\sigma$ and ends in a choice vertex $c_1$ or $c_2$.
Since $\sigma$ is winning, it plays the safe move of the table at this choice vertex.
In particular,
\begin{itemize}
\item $\nxt(c_1, \upd^*(\pi_2\rho_1)) = y$ and $\nxt(c_1, \upd^*(\pi_3\rho_1)) = b_4$;
\item $\nxt(c_2, \upd^*(\pi_1\rho_2)) = b_3$ and $\nxt(c_2, \upd^*(\pi_i\rho_2)) = b_2$ for $i = 2, 3$.
\end{itemize}
Since $\upd^*(\pi_i\rho_j) = \upd^*(\upd^*(\pi_i), \rho_j)$, the memory state at the end of $\pi_i\rho_j$ depends on $i$ only through the memory state~$\upd^*(\pi_i)$ at $x$.
So the first item gives $\upd^*(\pi_2) \neq \upd^*(\pi_3)$, and the second gives $\upd^*(\pi_1) \neq \upd^*(\pi_2)$ and $\upd^*(\pi_1) \neq \upd^*(\pi_3)$.
Hence, the memory states~$\upd^*(\pi_1)$, $\upd^*(\pi_2)$, $\upd^*(\pi_3)$ are pairwise different, and $|M| \geq 3$.
\end{proof}

\section{Five and Six Memory States}
\label{sec:cyclic}

Recall the question of Fijalkow and Horn~\cite{DBLP:journals/corr/abs-1010-2420} and of Bose et al.~\cite{DBLP:journals/corr/abs-2607-14199}: if \Adam wins a generalized reachability game with target sets of size at most two, does he also win it with four memory states?
The number four comes from an example of Löding with four colors, which both papers report they could not generalize to more.
The games~$\game_5$ and~$\game_6$ of this section are such generalizations and answer the question in the negative: they have five and six colors, and \Adam needs five and six memory states in them, respectively.
The idea is the one of Löding's example: \Adam has a single choice vertex~$c$, and \Eve has $k$ paths to $c$ that his memory must tell apart, just as it must tell her three paths to $x$ apart in $\game_3$.

Both games consist of two parts.
The first part is acyclic and leads from the initial vertex to a vertex~$c$: only \Eve moves in it, every path through it leaves at least one color unvisited, and \Eve chooses which.
The second part leads from $c$ back to $c$ and allows \Adam to avoid a single color: he chooses it at $c$, which is his only vertex.
Hence, \Adam wins if he chooses the same color at every visit of $c$, namely a color that \Eve left unvisited in the first part.
Thus, a corresponding memory structure requires one memory state per color.

We first present the game~$\game_5$ and prove that \Adam needs five memory states in it (Example~\ref{ex:m5} and Theorem~\ref{thm:m5}).
We then turn to the game~$\game_6$, in which he needs six (Example~\ref{ex:m6} and Theorem~\ref{thm:m6}).

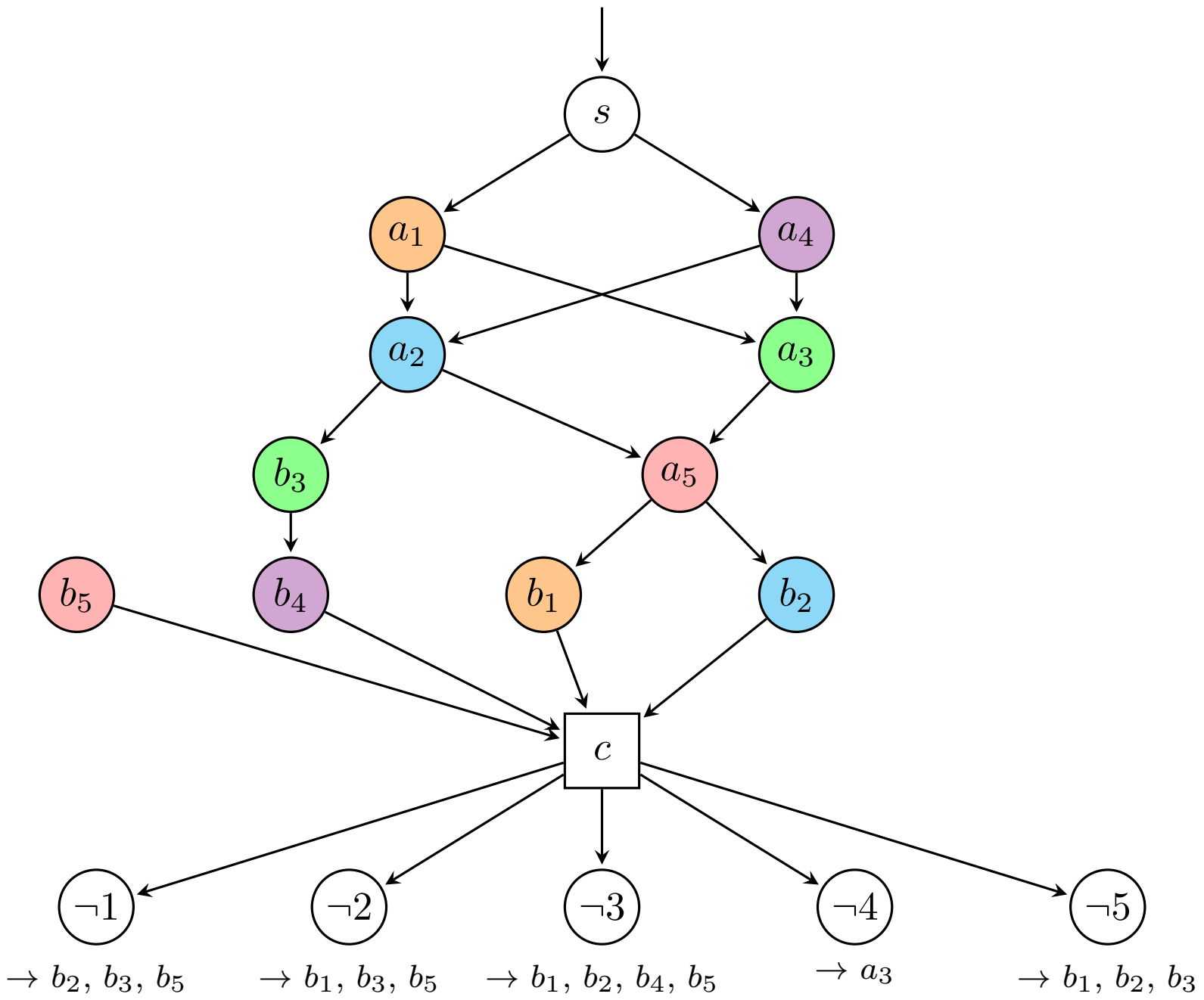
\begin{figure}[t]
\centering
\begin{tikzpicture}[game, x=1.7cm, y=1.05cm]
  \node[eve, initial above] (s)  at (2, 0)    {$s$};
  \node[eve, c1] (a1) at (1, -1)   {$a_1$};
  \node[eve, c4] (a4) at (3, -1)   {$a_4$};
  \node[eve, c2] (a2) at (1, -2)   {$a_2$};
  \node[eve, c3] (a3) at (3, -2)   {$a_3$};
  \node[eve, c3] (b3) at (0.4, -3) {$b_3$};
  \node[eve, c5] (a5) at (2.4, -3) {$a_5$};
  \node[eve, c5] (b5) at (-0.7, -4) {$b_5$};
  \node[eve, c4] (b4) at (0.4, -4) {$b_4$};
  \node[eve, c1] (b1) at (1.7, -4) {$b_1$};
  \node[eve, c2] (b2) at (3, -4)   {$b_2$};
  \node[adam] (c) at (2, -5.3) {$c$};
  \foreach \i/\x/\lab in {1/-0.6/{$b_2$, $b_3$, $b_5$}, 2/0.7/{$b_1$, $b_3$, $b_5$}, 3/2/{$b_1$, $b_2$, $b_4$, $b_5$}, 4/3.3/{$a_3$}, 5/4.6/{$b_1$, $b_2$, $b_3$}} {
    \node[eve] (o\i) at (\x, -6.6) {$\opt{\i}$};
    \node[font=\scriptsize, anchor=north] at (\x, -6.95) {$\to$ \lab};
  }
  \path[->]
    (s) edge (a1) (s) edge (a4)
    (a1) edge (a2) (a1) edge (a3) (a4) edge (a2) (a4) edge (a3)
    (a2) edge (b3) (a2) edge (a5) (a3) edge (a5)
    (b3) edge (b4) (a5) edge (b1) (a5) edge (b2)
    (b4) edge (c) (b1) edge (c) (b2) edge (c) (b5) edge (c)
    (c) edge (o1) (c) edge (o2) (c) edge (o3) (c) edge (o4) (c) edge (o5);
\end{tikzpicture}
\caption{The game~$\game_5$ of Example~\ref{ex:m5} ($k = 5$), with the target sets~$F_j = \set{a_j, b_j}$ for $j = 1, \dots, 5$.
The successors of each vertex~$\opt{i}$ are listed below it.}
\label{fig:m5}
\end{figure}

\begin{example}[\Adam needs five memory states; $k = 5$]
\label{ex:m5}
Figure~\ref{fig:m5} shows the game~$\game_5$ with initial vertex~$s$.
\Adam owns $c$, and \Eve owns all other vertices.
The target sets are $F_j = \set{a_j, b_j}$ for $j = 1, \dots, 5$.
The edges of $\game_5$ are drawn as arrows in Figure~\ref{fig:m5}, except that the successors of each vertex~$\opt{i}$ are listed below it.
Every cycle of $\game_5$ passes $c$.
\end{example}

The game proceeds in two stages.
In the first stage, \Eve chooses one of the ten paths from $s$ to the first visit of $c$ listed in the table below.
Each of them leaves at least one color unvisited.
Note that each path~$\pi_i$ with $i \leq 5$ leaves exactly one color unvisited, namely color~$i$.
Paths~$\pi_6, \ldots, \pi_{10}$, on the other hand, leave two colors unvisited.
\begin{center}
\begin{tabular}{@{}llll@{}}
\toprule
 & path & colors visited & colors not visited \\
\midrule
$\pi_1$ & $s\,a_4\,a_3\,a_5\,b_2\,c$ & $2, 3, 4, 5$ & $1$ \\
$\pi_2$ & $s\,a_4\,a_3\,a_5\,b_1\,c$ & $1, 3, 4, 5$ & $2$ \\
$\pi_3$ & $s\,a_4\,a_2\,a_5\,b_1\,c$ & $1, 2, 4, 5$ & $3$ \\
$\pi_4$ & $s\,a_1\,a_3\,a_5\,b_2\,c$ & $1, 2, 3, 5$ & $4$ \\
$\pi_5$ & $s\,a_1\,a_2\,b_3\,b_4\,c$ & $1, 2, 3, 4$ & $5$ \\
\midrule
$\pi_6$ & $s\,a_1\,a_2\,a_5\,b_1\,c$ & $1, 2, 5$ & $3, 4$ \\
$\pi_7$ & $s\,a_1\,a_2\,a_5\,b_2\,c$ & $1, 2, 5$ & $3, 4$ \\
$\pi_8$ & $s\,a_1\,a_3\,a_5\,b_1\,c$ & $1, 3, 5$ & $2, 4$ \\
$\pi_9$ & $s\,a_4\,a_2\,b_3\,b_4\,c$ & $2, 3, 4$ & $1, 5$ \\
$\pi_{10}$ & $s\,a_4\,a_2\,a_5\,b_2\,c$ & $2, 4, 5$ & $1, 3$ \\
\bottomrule
\end{tabular}
\end{center}

In the second stage, a play loops through $c$ infinitely often.
Following Fijalkow and Horn, we call the successors of $c$ the \emph{options}.
In $\game_5$, these are the uncolored vertices~$\opt{1}, \dots, \opt{5}$.
Furthermore, we say that a path that starts in an option, ends in $c$, and visits $c$ only once is a \emph{continuation} of this option.
The following table lists the continuations of the options of $\game_5$ (note that each row contains several continuations, separated by semicolons).
\begin{center}
\begin{tabular}{@{}lll@{}}
\toprule
option & continuations back to $c$ & colors \\
\midrule
$\opt{1}$ & $b_2$; $b_3 \to b_4$; $b_5$ & $2$; $3, 4$; $5$ \\
$\opt{2}$ & $b_1$; $b_3 \to b_4$; $b_5$ & $1$; $3, 4$; $5$ \\
$\opt{3}$ & $b_1$; $b_2$; $b_4$; $b_5$ & $1$; $2$; $4$; $5$ \\
$\opt{4}$ & $a_3 \to a_5 \to b_1$; $a_3 \to a_5 \to b_2$ & $3, 5, 1$; $3, 5, 2$ \\
$\opt{5}$ & $b_1$; $b_2$; $b_3 \to b_4$ & $1$; $2$; $3, 4$ \\
\bottomrule
\end{tabular}
\end{center}
We say that an option \emph{avoids} a color~$i$ if none of its continuations visits color~$i$.
Moreover, a vertex of color~$j$ on a continuation of an option is a \emph{threat of color~$j$} of this option.
By the table, the option~$\opt{i}$ avoids color~$i$ and has a threat of every other color.

The key idea of $\game_5$ is that \Adam has to move to the option~$\opt{i}$ after the path~$\pi_i$ with $i \leq 5$.
If \Adam moves to any other option, \Eve can move to its threat of color~$i$, visits all five colors, and wins.
If \Adam moves to $\opt{i}$ at every visit of $c$, \Eve cannot visit color~$i$ in any continuation of $\opt{i}$, and \Adam wins.
Hence, \Adam's memory has to tell the five paths~$\pi_1, \dots, \pi_5$ apart, resulting in five memory states.
This idea is formalized in the proof of Theorem~\ref{thm:m5} below.

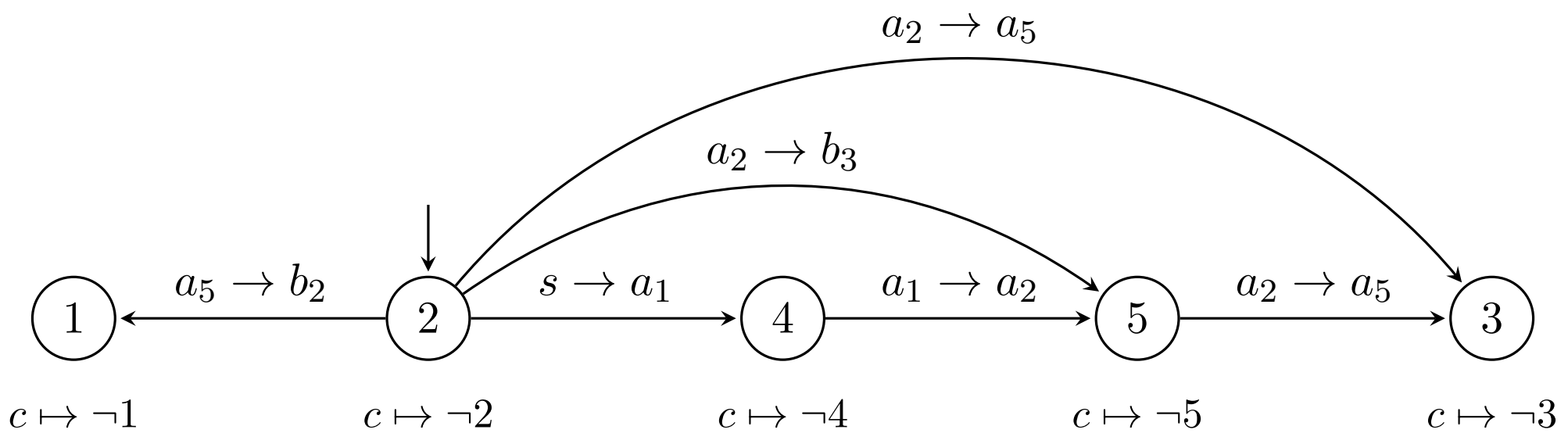
\begin{figure}[ht]
\centering
\begin{tikzpicture}[->, semithick, initial text=, initial distance=6mm,
                    every state/.style={minimum size=7mm, inner sep=1pt}]
  \node[state]                (q1) at (-3, 0) {$1$};
  \node[state, initial above] (q2) at (0, 0)  {$2$};
  \node[state]                (q4) at (3, 0)  {$4$};
  \node[state]                (q5) at (6, 0)  {$5$};
  \node[state]                (q3) at (9, 0)  {$3$};
  \path (q2) edge node[above] {$a_5 \to b_2$} (q1)
        (q2) edge node[above] {$s \to a_1$}   (q4)
        (q4) edge node[above] {$a_1 \to a_2$} (q5)
        (q5) edge node[above] {$a_2 \to a_5$} (q3)
        (q2) edge[bend left=35] node[above] {$a_2 \to b_3$} (q5)
        (q2) edge[bend left=50] node[above] {$a_2 \to a_5$} (q3);
  \foreach \q in {1, 2, 3, 4, 5}
    \node[below=2mm of q\q, font=\small] {$c \mapsto \opt{\q}$};
\end{tikzpicture}
\caption{A five-state winning strategy of \Adam in $\game_5$, in the notation of Figure~\ref{fig:m3-strategy}.}
  \label{fig:m5-strategy}
\end{figure}

\begin{theorem}
\label{thm:m5}
\Adam wins $\game_5$ with five memory states, and every winning strategy of \Adam has at least five (i.e., $\memof(\game_5) = 5$).
\end{theorem}

\begin{proof}[Proof of Theorem~\ref{thm:m5}]
We prove the two claims of the theorem in turn.
First, we give a strategy of \Adam with five memory states and show that it is winning.
Second, we show that every winning strategy of \Adam requires at least five memory states.

\medskip
\noindent\textit{Five states suffice.}
Figure~\ref{fig:m5-strategy} shows a winning strategy of \Adam with five memory states.
To simplify our notation, we identify the memory states with the numbers $1, 2, 3, 4, 5$.
Transitions not drawn in the figure leave the memory state unchanged.

It is not hard to verify that at the end of the path $\pi_i$ with $i \leq 5$, the memory is in state~$i$.
Furthermore, at the end of the paths $\pi_6$, $\pi_7$ and $\pi_{10}$ it is in state~$3$, at the end of $\pi_8$ in state~$4$, and at the end of $\pi_9$ in state~$5$.
In each case, the memory state at the end of the path is a color that the path leaves unvisited.

Now consider a play that is consistent with this strategy, and let $q$ be the memory state at its first visit of $c$.
Up to this visit, the play follows one of the paths~$\pi_1, \dots, \pi_{10}$, which leaves color~$q$ unvisited.
After this visit, the play stays among $c$, the options, $a_3$, $a_5$ and the vertices~$b_1, \dots, b_5$.
Among the edges between these vertices, only $a_5 \to b_2$ can change the memory, and only in state~$2$.
There, the play enters $a_5$ only along $c\,\opt{4}\,a_3\,a_5$.
\Adam moves from $c$ to $\opt{4}$ only in memory state~$4$, and the edges~$\opt{4} \to a_3$ and $a_3 \to a_5$ keep this state, so the play reaches $a_5$ in state~$4$.
Hence, the edge~$a_5 \to b_2$ never changes the memory after the first visit of $c$, and the memory stays in state~$q$.

Therefore, \Adam's strategy moves to $\opt{q}$ at every visit of $c$.
Since the path to the first visit of $c$ leaves color~$q$ unvisited and $\opt{q}$ avoids color~$q$, the play never visits color~$q$.
Thus, the strategy is winning.

\medskip
\noindent\textit{Four states do not suffice.}
Let $\sigma = (\mem, \nxt)$ with $\mem = (M, q_0, \upd)$ be a winning strategy of \Adam, and let $i \in \set{1, \dots, 5}$.
The path~$\pi_i$ contains no \Adam vertex before $c$, so it is a finite play consistent with $\sigma$.
Moreover, the memory state at its end is $\upd^*(\pi_i)$, the update along the edge into $c$ included.

Let now $j \neq i$ and assume that $\nxt(c, \upd^*(\pi_i)) = \opt{j}$.
By the table, the option~$\opt{j}$ has a threat of color~$i$.
Let $\rho$ be a continuation of $\opt{j}$ through this threat.
Since all vertices of $\rho$ before $c$ belong to \Eve, the finite play that follows $\pi_i$ and then $\rho$ is consistent with $\sigma$.
This finite play visits all colors: color~$i$ on $\rho$ and every other color on $\pi_i$.
This contradicts the assumption that $\sigma$ is winning, so $\nxt(c, \upd^*(\pi_i)) \neq \opt{j}$.
Hence, $\nxt(c, \upd^*(\pi_i)) = \opt{i}$ for every $i$, the memory states~$\upd^*(\pi_1), \dots, \upd^*(\pi_5)$ are pairwise different, and $|M| \geq 5$.
\end{proof}

With one more color, the same mechanism gives six memory states.

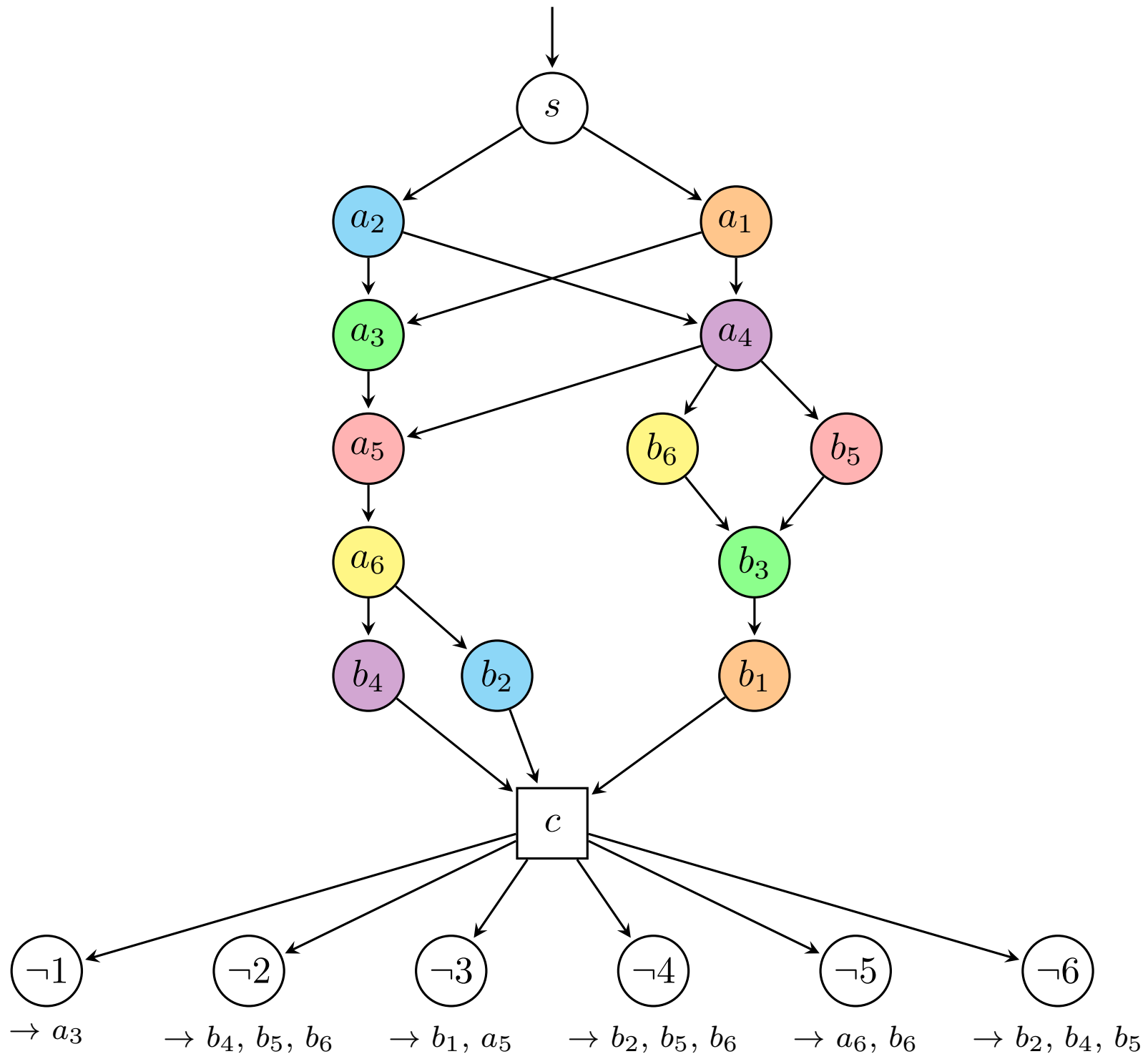
\begin{figure}[t]
\centering
\begin{tikzpicture}[game, x=1.7cm, y=1.05cm]
  \node[eve, initial above] (s)  at (2, 0)    {$s$};
  \node[eve, c2] (a2) at (1, -1)   {$a_2$};
  \node[eve, c1] (a1) at (3, -1)   {$a_1$};
  \node[eve, c3] (a3) at (1, -2)   {$a_3$};
  \node[eve, c4] (a4) at (3, -2)   {$a_4$};
  \node[eve, c5] (a5) at (1, -3)   {$a_5$};
  \node[eve, c6] (b6) at (2.6, -3) {$b_6$};
  \node[eve, c5] (b5) at (3.6, -3) {$b_5$};
  \node[eve, c6] (a6) at (1, -4)   {$a_6$};
  \node[eve, c3] (b3) at (3.1, -4) {$b_3$};
  \node[eve, c4] (b4) at (1.0, -5) {$b_4$};
  \node[eve, c2] (b2) at (1.7, -5) {$b_2$};
  \node[eve, c1] (b1) at (3.1, -5) {$b_1$};
  \node[adam] (c) at (2, -6.3) {$c$};
  \foreach \i/\x/\lab in {1/-0.75/{$a_3$}, 2/0.35/{$b_4$, $b_5$, $b_6$}, 3/1.45/{$b_1$, $a_5$}, 4/2.55/{$b_2$, $b_5$, $b_6$}, 5/3.65/{$a_6$, $b_6$}, 6/4.75/{$b_2$, $b_4$, $b_5$}} {
    \node[eve] (o\i) at (\x, -7.6) {$\opt{\i}$};
    \node[font=\scriptsize, anchor=north] at (\x, -7.95) {$\to$ \lab};
  }
  \path[->]
    (s) edge (a2) (s) edge (a1)
    (a2) edge (a3) (a2) edge (a4) (a1) edge (a3) (a1) edge (a4)
    (a3) edge (a5) (a4) edge (a5) (a4) edge (b6) (a4) edge (b5)
    (a5) edge (a6) (b6) edge (b3) (b5) edge (b3)
    (a6) edge (b4) (a6) edge (b2) (b3) edge (b1)
    (b4) edge (c) (b2) edge (c) (b1) edge (c)
    (c) edge (o1) (c) edge (o2) (c) edge (o3) (c) edge (o4) (c) edge (o5) (c) edge (o6);
\end{tikzpicture}
\caption{The game~$\game_6$ of Example~\ref{ex:m6} ($k = 6$), with the target sets~$F_j = \set{a_j, b_j}$ for $j = 1, \dots, 6$.
The successors of each vertex~$\opt{i}$ are listed below it.}
\label{fig:m6}
\end{figure}

\begin{example}[\Adam needs six memory states; $k = 6$]
\label{ex:m6}
Figure~\ref{fig:m6} shows the game~$\game_6$ with initial vertex~$s$.
\Adam owns $c$, and \Eve owns all other vertices.
The target sets are $F_j = \set{a_j, b_j}$ for $j = 1, \dots, 6$.
The edges of $\game_6$ are drawn as arrows in Figure~\ref{fig:m6}, except that the successors of each vertex~$\opt{i}$ are listed below it.
Every cycle of $\game_6$ passes $c$.
\end{example}

The game~$\game_6$ proceeds in the same two stages as $\game_5$.
In the first stage, \Eve chooses one of the twelve paths from $s$ to the first visit of $c$ listed in the table below.
Again, each path~$\pi_i$ with $i \leq 6$ leaves exactly one color unvisited, namely color~$i$.
Paths~$\pi_7, \ldots, \pi_{12}$ leave two colors unvisited.
\begin{center}
\begin{tabular}{@{}llll@{}}
\toprule
 & path & colors visited & colors not visited \\
\midrule
$\pi_1$ & $s\,a_2\,a_3\,a_5\,a_6\,b_4\,c$ & $2, 3, 4, 5, 6$ & $1$ \\
$\pi_2$ & $s\,a_1\,a_3\,a_5\,a_6\,b_4\,c$ & $1, 3, 4, 5, 6$ & $2$ \\
$\pi_3$ & $s\,a_1\,a_4\,a_5\,a_6\,b_2\,c$ & $1, 2, 4, 5, 6$ & $3$ \\
$\pi_4$ & $s\,a_1\,a_3\,a_5\,a_6\,b_2\,c$ & $1, 2, 3, 5, 6$ & $4$ \\
$\pi_5$ & $s\,a_2\,a_4\,b_6\,b_3\,b_1\,c$ & $1, 2, 3, 4, 6$ & $5$ \\
$\pi_6$ & $s\,a_2\,a_4\,b_5\,b_3\,b_1\,c$ & $1, 2, 3, 4, 5$ & $6$ \\
\midrule
$\pi_7$ & $s\,a_1\,a_4\,a_5\,a_6\,b_4\,c$ & $1, 4, 5, 6$ & $2, 3$ \\
$\pi_8$ & $s\,a_2\,a_4\,a_5\,a_6\,b_4\,c$ & $2, 4, 5, 6$ & $1, 3$ \\
$\pi_9$ & $s\,a_2\,a_4\,a_5\,a_6\,b_2\,c$ & $2, 4, 5, 6$ & $1, 3$ \\
$\pi_{10}$ & $s\,a_1\,a_4\,b_6\,b_3\,b_1\,c$ & $1, 3, 4, 6$ & $2, 5$ \\
$\pi_{11}$ & $s\,a_1\,a_4\,b_5\,b_3\,b_1\,c$ & $1, 3, 4, 5$ & $2, 6$ \\
$\pi_{12}$ & $s\,a_2\,a_3\,a_5\,a_6\,b_2\,c$ & $2, 3, 5, 6$ & $1, 4$ \\
\bottomrule
\end{tabular}
\end{center}

In the second stage, a play again loops through $c$ infinitely often.
The following table lists the continuations of the options~$\opt{1}, \dots, \opt{6}$ of $\game_6$.
\begin{center}
\begin{tabular}{@{}lll@{}}
\toprule
option & continuations back to $c$ & colors \\
\midrule
$\opt{1}$ & $a_3 \to a_5 \to a_6 \to b_4$; $a_3 \to a_5 \to a_6 \to b_2$ & $3, 5, 6, 4$; $3, 5, 6, 2$ \\
$\opt{2}$ & $b_6 \to b_3 \to b_1$; $b_5 \to b_3 \to b_1$; $b_4$ & $6, 3, 1$; $5, 3, 1$; $4$ \\
$\opt{3}$ & $a_5 \to a_6 \to b_4$; $a_5 \to a_6 \to b_2$; $b_1$ & $5, 6, 4$; $5, 6, 2$; $1$ \\
$\opt{4}$ & $b_6 \to b_3 \to b_1$; $b_5 \to b_3 \to b_1$; $b_2$ & $6, 3, 1$; $5, 3, 1$; $2$ \\
$\opt{5}$ & $a_6 \to b_4$; $a_6 \to b_2$; $b_6 \to b_3 \to b_1$ & $6, 4$; $6, 2$; $6, 3, 1$ \\
$\opt{6}$ & $b_5 \to b_3 \to b_1$; $b_4$; $b_2$ & $5, 3, 1$; $4$; $2$ \\
\bottomrule
\end{tabular}
\end{center}
As in $\game_5$, the option~$\opt{i}$ avoids color~$i$ and has a threat of every other color.
Hence, the key idea of $\game_5$ carries over: \Adam has to move to the option~$\opt{i}$ after the path~$\pi_i$ with $i \leq 6$, and his memory has to tell the six paths~$\pi_1, \dots, \pi_6$ apart.

\begin{figure}[ht]
\centering
\begin{tikzpicture}[->, semithick, initial text=, initial distance=5mm,
                    every state/.style={minimum size=7mm, inner sep=1pt}]
  \node[state]                (q4) at (0, 0)    {$4$};
  \node[state]                (q2) at (2.5, 0)  {$2$};
  \node[state, initial above] (q1) at (5, 0)    {$1$};
  \node[state]                (q3) at (7.5, 0)  {$3$};
  \node[state]                (q5) at (10, 0)   {$5$};
  \node[state]                (q6) at (12.5, 0) {$6$};
  \path (q1) edge node[above] {$s \to a_1$}   (q2)
        (q2) edge node[above] {$a_6 \to b_2$} (q4)
        (q1) edge node[above] {$a_2 \to a_4$} (q3)
        (q3) edge node[above] {$a_4 \to b_6$} (q5)
        (q2) edge[bend left=60] node[above] {$a_1 \to a_4$} (q3)
        (q3) edge[bend left=50] node[above] {$a_4 \to b_5$} (q6);
  \foreach \q in {1, 2, 3, 4, 5, 6}
    \node[below=2mm of q\q, font=\small] {$c \mapsto \opt{\q}$};
\end{tikzpicture}
\caption{A six-state winning strategy of \Adam in $\game_6$, in the notation of Figure~\ref{fig:m3-strategy}.}
  \label{fig:m6-strategy}
\end{figure}
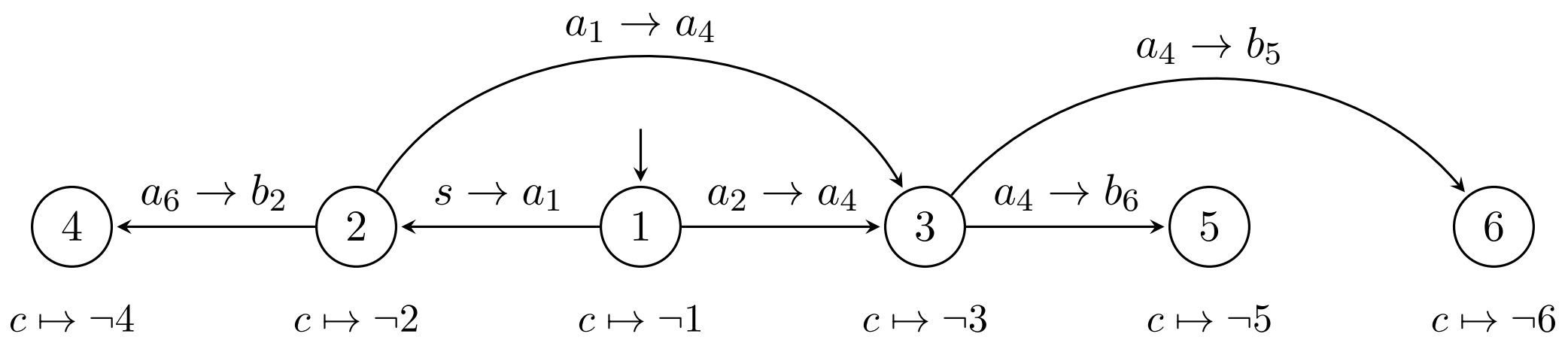

\begin{theorem}
\label{thm:m6}
\Adam wins $\game_6$ with six memory states, and every winning strategy of \Adam has at least six (i.e., $\memof(\game_6) = 6$).
\end{theorem}

\begin{proof}[Proof of Theorem~\ref{thm:m6}]
The proof follows the one of Theorem~\ref{thm:m5}, and we only point out the differences.

\medskip
\noindent\textit{Six states suffice.}
Figure~\ref{fig:m6-strategy} shows a winning strategy of \Adam with six memory states, which we name $1, \dots, 6$.
It is not hard to verify that at the end of the path $\pi_i$ with $i \leq 6$, the memory is in state~$i$.
Furthermore, at the end of the paths $\pi_7$, $\pi_8$ and $\pi_9$ it is in state~$3$, at the end of $\pi_{10}$ in state~$5$, at the end of $\pi_{11}$ in state~$6$, and at the end of $\pi_{12}$ in state~$1$.
In each case, the memory state at the end of the path is a color that the path leaves unvisited.

After its first visit of $c$, a play no longer visits $s$, $a_1$, $a_2$ and $a_4$.
Hence, the only edge that can then change the memory is $a_6 \to b_2$, and only in state~$2$.
In $\game_5$, the corresponding vertex~$a_5$ is entered only after the option~$\opt{4}$.
In $\game_6$, the vertex~$a_6$ is entered only after one of the options~$\opt{1}$, $\opt{3}$ and $\opt{5}$, along $c\,\opt{1}\,a_3\,a_5\,a_6$, $c\,\opt{3}\,a_5\,a_6$, and $c\,\opt{5}\,a_6$, respectively.
\Adam moves to these options only in the memory states~$1$, $3$ and $5$, and the edges on the way to $a_6$ keep this state.
So the play never reaches $a_6$ in state~$2$.
Hence, the edge~$a_6 \to b_2$ never changes the memory after the first visit of $c$, and the memory stays in the state it has at this visit.
Therefore, as in the proof of Theorem~\ref{thm:m5}, the strategy is winning.

\medskip
\noindent\textit{Five states do not suffice.}
The argument of Theorem~\ref{thm:m5} applies to the paths~$\pi_1, \dots, \pi_6$.
Each path~$\pi_i$ contains no \Adam vertex before $c$, and every option~$\opt{j}$ with $j \neq i$ has a threat of color~$i$.
Hence, every winning strategy of \Adam has at least six memory states.
\end{proof}

Let us conclude this section with a remark on the relation of our games~$\game_5$ and~$\game_6$ to Löding's example.

\begin{remark}
\label{rem:loding}
The games~$\game_5$ and~$\game_6$ have the shape of Löding's example~\cite[Fig.~5]{DBLP:journals/corr/abs-1010-2420}: \Eve alone moves before \Adam's only choice vertex~$c$, where he has one option per color, which avoids this color and has a threat of every other color.
In Löding's example, every path from $s$ that visits $c$ only at its end has exactly one vertex before $c$ on a cycle through~$c$, namely its last colored vertex.
Furthermore, every option leads directly to such vertices.
By contrast, in $\game_5$ and $\game_6$, some options lead into the middle of these paths.
In $\game_5$, the path~$\pi_5$ ends with $b_3\,b_4$, and $b_3$ leads only to $b_4$.
So the option~$\opt{4}$, which avoids color~$4$, cannot use $b_3$ as its threat of color~$3$.
It leads to $a_3$ instead, which lies in the middle of $\pi_1$, $\pi_2$ and~$\pi_4$.
In $\game_6$, the paths~$\pi_5$ and~$\pi_6$ end with $b_6\,b_3\,b_1$ and~$b_5\,b_3\,b_1$.
For analogous reasons, the options~$\opt{1}$ and~$\opt{3}$ lead to $a_3$ and~$a_5$, respectively.
\end{remark}

\section{Five or More Target Sets}
\label{sec:five-or-more}

Recall the two conjectured bounds on \Adam's memory in games with target sets of size at most two: four memory states on every arena~\cite{DBLP:journals/corr/abs-1010-2420,DBLP:journals/corr/abs-2607-14199} and two on acyclic arenas~\cite{FijalkowBlog2018}.
Both concern every number~$k$ of colors, whereas the games of Sections~\ref{sec:three} and~\ref{sec:cyclic} have five or six colors.
We now show that both bounds fail for every $k \geq 5$.
For this, it suffices that a lower bound on \Adam's memory for $k$ colors is also one for $k + 1$ colors, which holds on both acyclic and cyclic arenas.
We obtain this result by adding a new color whose two vertices every play visits before reaching the initial vertex.

\begin{lemma}[Padding]
\label{lem:padding}
Let $\game = (\arena, s, F_1, \dots, F_k)$ be a generalized reachability game that \Adam wins.
Let $\game'$ be the game obtained from $\game$ by adding two new vertices~$a_{k+1}$ and~$b_{k+1}$ of \Eve, the edges~$a_{k+1} \to b_{k+1}$ and~$b_{k+1} \to s$, and the target set~$F_{k+1} = \set{a_{k+1}, b_{k+1}}$, with the initial vertex~$a_{k+1}$.
Then \Adam wins $\game'$ and $\memof(\game') = \memof(\game)$.
Moreover, if $\game$ is acyclic, then so is $\game'$.
\end{lemma}

\begin{proof}
No edge leads into $a_{k+1}$, and the only edge into $b_{k+1}$ comes from $a_{k+1}$.
Hence, neither of the two vertices lies on a cycle, which shows the last claim.
Moreover, the plays of $\game'$ are the sequences~$a_{k+1}\,b_{k+1}\,\rho$ with a play~$\rho$ of $\game$.
Every such play visits color~$k+1$.
So \Adam wins $a_{k+1}\,b_{k+1}\,\rho$ in $\game'$ if and only if he wins $\rho$ in $\game$.

Let $(\mem, \nxt)$ with $\mem = (M, q_0, \upd)$ be a winning strategy of \Adam in $\game$ with $|M| = \memof(\game)$.
Extend $\upd$ to the two new edges such that both keep the memory state.
Then the memory state at the finite play~$a_{k+1}\,b_{k+1}\,\pi$ of $\game'$ is $\upd^*(\pi)$ for every finite play~$\pi$ of $\game$.
Since \Adam owns neither new vertex, a play~$a_{k+1}\,b_{k+1}\,\rho$ is consistent with the extended strategy if and only if $\rho$ is consistent with $(\mem, \nxt)$.
Since $(\mem, \nxt)$ is winning, \Adam wins every such $\rho$ and hence every play of $\game'$ consistent with the extended strategy.
Thus, \Adam wins $\game'$ and $\memof(\game') \leq \memof(\game)$.

Conversely, let $(\mem', \nxt')$ with $\mem' = (M, q_0', \upd')$ be a winning strategy of \Adam in $\game'$ with $|M| = \memof(\game')$.
Let $\mem = (M, q_0, \upd)$, where $q_0 = \upd'^*(q_0', a_{k+1}\,b_{k+1}\,s)$ and $\upd$ is the restriction of $\upd'$ to the edges of $\game$.
Since the update is applied edge by edge, $\upd'^*(q_0', a_{k+1}\,b_{k+1}\,\pi) = \upd^*(q_0, \pi)$ for every finite play~$\pi$ of $\game$.
Hence, a play~$\rho$ of $\game$ is consistent with $(\mem, \nxt')$ if and only if $a_{k+1}\,b_{k+1}\,\rho$ is consistent with $(\mem', \nxt')$.
Since $(\mem', \nxt')$ is winning, \Adam wins every such play~$a_{k+1}\,b_{k+1}\,\rho$ and hence every play~$\rho$ of $\game$ consistent with $(\mem, \nxt')$.
Thus, $\memof(\game) \leq \memof(\game')$.
\end{proof}

Applying Lemma~\ref{lem:padding} repeatedly to $\game_3$ (Theorem~\ref{thm:m3}) now yields, for every $k \geq 5$, an acyclic game with $k$ target sets of size two in which \Adam needs exactly three memory states.
Thus, the bound of two memory states on acyclic arenas fails for every $k \geq 5$.
Applying the lemma repeatedly to $\game_5$ and $\game_6$ (Theorems~\ref{thm:m5} and~\ref{thm:m6}) yields games with $k$ target sets of size two in which \Adam needs exactly five and six memory states for every $k \geq 5$ and $k \geq 6$, respectively.
Hence, the bound of four memory states fails for every $k \geq 5$.

\section{Conclusion}
\label{sec:conclusion}

We have refuted both conjectured bounds on \Adam's memory in generalized reachability games with target sets of size at most two: four memory states on every arena and two on acyclic arenas.
In the acyclic game~$\game_3$, \Adam needs three memory states instead of two.
In the cyclic games~$\game_5$ and~$\game_6$, he needs five and six memory states, respectively, instead of four.
By Lemma~\ref{lem:padding}, both bounds fail for every $k \geq 5$.

All three games rest on one mechanism.
Before \Adam moves, \Eve chooses among several paths that lead to a common vertex but visit different sets of colors.
His good moves later in the play depend on the colors that the chosen path visits.
Hence, his memory has to tell these paths apart.

Several questions remain open.
The four-state bound is open for $k = 4$, and the two-state bound on acyclic arenas for $k = 3$ and $k = 4$.
For every $k \geq 6$, the largest known lower bound on \Adam's memory is six, from~$\game_6$.
Whether his memory grows with~$k$ is not known, even on acyclic arenas.
A bound independent of~$k$ would still place the problem of deciding the winner in~$\mathrm{coNP}^{\mathrm{NP}}$.
The exact complexity of this problem remains the main open question.

\paragraph{AI statement.}
The author used generative AI (Claude, Anthropic) in this work.

\bibliographystyle{plain}
\bibliography{refs}

\end{document}